\documentclass[prd,twocolumn,preprintnumbers,amsmath,amssymb,superscriptaddress,showpacs,showkeys,nofootinbib]{revtex4-1}
\pdfoutput=1
\usepackage{amsmath,amsfonts,amssymb,amsthm,amstext,amscd,eucal,srcltx,mathrsfs}
\usepackage{epsfig,graphicx,bm}
\usepackage{epstopdf, epsf}
\usepackage{dcolumn}
\usepackage{hyperref}
\usepackage{tikz}
\usetikzlibrary{arrows,shapes}
\usetikzlibrary{matrix,arrows}
\newcommand{\be}{\begin{eqnarray}}
\newcommand{\ee}{\end{eqnarray}}

\renewcommand{\d}{\mbox{${\rm d}$}} 
\newcommand{\lp}{\ell_{\rm p}}
\newcommand{\mpl}{m_{\rm p}}
\newcommand{\gn}{G_{\rm N}}

\begin{document}
\title{Corpuscular slow-roll inflation}
\author{Roberto Casadio}
\email{casadio@bo.infn.it}
\affiliation{Dipartimento di Fisica e Astronomia,
Alma Mater Universit\`a di Bologna,
via~Irnerio~46, 40126~Bologna, Italy}
\affiliation{I.N.F.N., Sezione di Bologna, IS FLAG
viale~B.~Pichat~6/2, I-40127 Bologna, Italy}
\author{Andrea Giugno}
\email{A.Giugno@physik.uni-muenchen.de}
\affiliation{Arnold Sommerfeld Center, Ludwig-Maximilians-Universit\"at,
Theresienstra{\ss}e 37, 80333 M\"unchen, Germany}
\author{Andrea Giusti}
\email{andrea.giusti@bo.infn.it}
\affiliation{Dipartimento di Fisica e Astronomia,
Alma Mater Universit\`a di Bologna,
via~Irnerio~46, 40126~Bologna, Italy}
\affiliation{I.N.F.N., Sezione di Bologna, IS FLAG,
viale~B.~Pichat~6/2, I-40127 Bologna, Italy}
\affiliation{Arnold Sommerfeld Center, Ludwig-Maximilians-Universit\"at,
Theresienstra{\ss}e 37, 80333 M\"unchen, Germany}
\begin{abstract}
We show that a corpuscular description of gravity can lead to an inflationary scenario similar
to Starobinsky's model without requiring the introduction of the inflaton field.
All relevant properties are determined by the number of gravitons in the cosmological
condensate or, equivalently, by their Compton length.
In particular, the relation between the Hubble parameter $H$ and its time derivative $\dot H$
required by CMB observations at the end of inflation, as well as the
(minimum) initial value of the slow-roll parameter, are naturally obtained from the Compton size
of the condensate.
\end{abstract}
\maketitle
\section{Introduction}
\label{intro}
The inflationary scenario in cosmology was introduced by Starobinsky~\cite{starobinskyI} and Guth~\cite{Guth-first} in the
early eighties in order to explain the homogeneity and flatness of our universe.
The new inflationary scenario was later proposed by Linde~\cite{Linde}, and Albrecht and Steinhardt~\cite{AS}, in which 
the accelerated expansion was driven by a scalar field (the {\em inflaton\/}) slowly rolling down a plateau of the potential
toward the minimum. 
If the plateau is sufficiently flat, the process lasts long enough to solve the cosmological problems mentioned above.
Moreover, the inflationary model based on the inflaton can be formally mapped into a $f(R)$ (modified) theory of gravity
(see e.g.~\cite{starobinsky}).
Nowadays, this scenario has become, almost unanimously, accepted as part of the standard model of the cosmo
and one case that appears particularly favoured by present observations~\cite{planck,WMAP} is precisely
Starobinsky's model~\cite{starobinskyI}.
\par	
Most models of inflation make use of the semiclassical approximation, in which the (background) metric is classical.
However, we are not guaranteed that this approximation is not missing relevant quantum properties of gravity in the early
universe~\cite{dvali}.
In this regard, the classical geometry of space-time could as well be conceived as an emerging property of a
coherent state describing a large number of gravitons, in close analogy to photons in a laser beam.
A peculiar feature of gravity is the attractive graviton-graviton interaction, which allows for their collapse and formation
of Bose-Einstein condensates. 
In Ref.~\cite{Gia1}, it was conjectured that this picture can reliably describe the physics inside a black hole,
which is in turn considered a compact quantum system on the verge of a phase transition.
Even when the gravitational regime is strong, the set-up is nicely understood as a Newtonian
theory of $N$ gravitons, which are loosely confined in a ``potential well" of size their Compton
wavelength $\lambda$ and interact with an effective gravitational coupling $\alpha\sim 1/N$.
As a result, it is possible to recover the correct post-Newtonian expansion of the gravitational field
generated by a static, spherically symmetric source~\cite{baryons} or the renowned
Bekenstein-Hawking area law~\cite{bekenstein} with logarithmic corrections~\cite{planckian} for the Hawking
radiation.
On the other hand, this framework represents a natural scenario for a cosmological model of
inflation~\cite{dvali,cko}, whose characteristic quantities display quantum properties related to the
corpuscular nature of gravity.
It will also appear that this description can help to constrain possible modified metric theories of
gravity~\cite{DeFelice:2010aj}, therefore proving to be an interesting benchmark.
\par
In this work we shall show that the corpuscular description of gravity can reproduce the inflationary expansion, 
purely as a consequence of the graviton self-interaction.
Unlike what was considered in Refs.~\cite{dvali,cko}, the primordial cosmological condensate can give rise to the
dynamics of Starobinsky's model~\cite{starobinskyI}, without requiring the introduction of an inflaton.
\section{Corpuscular cosmology}
\label{sez:cc}
Let us start from the assumption that matter and the corpuscular state of gravitons together
must reproduce the Friedmann equation of cosmology, which we write as the Hamiltonian
constraint
\be
\mathcal{H}_{\rm M}+\mathcal{H}_{\rm G}
=
0
\ ,
\label{H0}
\ee
where $\mathcal{H}_{\rm M}$ is the matter energy and $\mathcal{H}_{\rm G}$ the analogue quantity
for the graviton state.
We recall that local (Newton or Einstein) gravity being attractive in general implies that $\mathcal{H}_{\rm G}\le 0$,
although this is not true for the graviton self-interaction~\cite{baryons}, and might not be true
for the cosmological condensate of gravitons as a whole, as we are now going to discuss.
\subsection{Corpuscular De~Sitter}
\label{sub:corpDS}
In order to obtain the de~Sitter space-time in general relativity, one must assume the existence of a
cosmological constant term, or vacuum energy density $\rho_\Lambda$, so that the Friedmann equation
reads~\footnote{We shall use unites with $c=1$ and the Newton constant $\gn=\lp/\mpl$, where $\lp$
and $\mpl$ are the Planck length and mass respectively, and $\hbar=\lp\,\mpl$.}
\be
H^2
\equiv
\left(\frac{\dot a}{a}\right)^2
=
\frac{8}{3}\,\pi\,\gn\,\rho_\Lambda
\ .
\label{frwDS}
\ee
Upon integrating on the volume inside the Hubble radius that solves Eq.~\eqref{frwDS}, that is
$L_\Lambda=H^{-1}_\Lambda$, we obtain~\footnote{Factors of order one will be often omitted
from now on.}
\be
L_\Lambda 
\simeq
\gn\,L_\Lambda^3\,\rho_\Lambda
\simeq
\lp\,\frac{M_\Lambda}{\mpl}
\ ,
\label{LgM}
\ee
which looks exactly like the expression of the horizon radius for a black hole of ADM mass
$M_\Lambda$, and is the reason it was conjectured that the de~Sitter space-time could likewise
be viewed as a condensate of gravitons~\cite{dvali}.
\par
One can roughly describe the corpuscular model on assuming that the (soft virtual) graviton
self-interaction gives rise to a condensate of $N_\Lambda$ gravitons of typical Compton length
$\lambda\simeq L_\Lambda$, so that $M_\Lambda=N_\Lambda\,{\lp\,\mpl}/{L_\Lambda}$,
and the usual consistency condition
\be
M_\Lambda
\sim
\sqrt{N_\Lambda}\,\mpl
\label{MN}
\ee
for the graviton condensate immediately follows from Eq.~\eqref{LgM}.
Equivalently, one finds
\be
L_\Lambda
\sim
\sqrt{N_\Lambda}\,\lp
\ ,
\label{LMn}
\ee
which shows that for a macroscopic universe one needs $N_\Lambda\gg 1$.
Note also that we have
$\rho_\Lambda\sim L_\Lambda^{-3}\,{M_\Lambda}\sim {1}/{N_\Lambda}$,
so that the number of gravitons in the vacuum increases for smaller vacuum energy,
and $L_\Lambda \sim M_\Lambda \sim {1}/{\sqrt{\rho_\Lambda}}$.
It is important to remark that the above relations do not need to hold for gravitons that are
not in the condensate, therefore one expects deviations occur if regular matter is added~\cite{Cadoni:2017evg},
or if the system is driven out of equilibrium.
%
%
%
%
%
\par
We can refine the above corpuscular description of the de~Sitter space by following the line of reasoning of
Refs.~\cite{baryons}, where it was shown that the maximal packing condition which yields the
scaling relations~\eqref{LMn} for a black hole actually follows from the energy balance~\eqref{H0}
when matter becomes totally negligible.
In the present case, matter is absent {\em a priori\/} and $\mathcal{H}_{\rm M}=0$, so that one is left with
\be
\mathcal{H}_{\rm G}
\simeq
U_{\rm N}
+
U_{\rm PN}
=
0
\ .
\label{H00}
\ee
The negative ``Newtonian energy'' of the $N_\Lambda$ gravitons can be obtained 
from a coherent state description of the condensate~\cite{baryons} in which each graviton has negative
binding energy $\varepsilon_\Lambda$ given by the Compton relation, that is
\be
U_{\rm N}
\simeq
M_\Lambda\,\phi_{\rm N}
=
N_\Lambda\,\varepsilon_{\Lambda}
=
-N_\Lambda\,\frac{\lp\,\mpl}{L_\Lambda}
\ .
\label{UN0}
\ee
The positive ``post-Newtonian'' contribution is then given by the graviton self-interaction term~\cite{baryons}
\be
U_{\rm PN}
\simeq
N_\Lambda\,\varepsilon_\Lambda\,\phi_{\rm N}
=
N_\Lambda^{3/2}\,\frac{\lp^2\,\mpl}{L_\Lambda^2}
\ ,
\label{UPN0}
\ee
where we used the Newtonian potential 
\be
\phi_{\rm N}
=
-\frac{N_\Lambda\,\lp\,\mpl}{M_\Lambda\,L_\Lambda}
=
-\sqrt{N_\Lambda}\,\frac{\lp}{L_\Lambda}
\ ,
\ee
as follows from Eq.~\eqref{UN0} and the scaling relation~\eqref{MN}.
\subsection{Metric de~Sitter}
\label{sub:metricDS}
Before we consider explicit ways of perturbing the de~Sitter solution,
let us try to re-interpret our results in terms of a metric theory.
We have just seen that the de~Sitter universe is, in a sense, a solution of our Hamiltonian constraint
and we also know the de~Sitter metric is an exact solution of a modified theory of gravity~\cite{DeFelice:2010aj}
\be
S
=
\frac{1}{16\,\pi\,\gn}
\int \d^4 x\,\sqrt{-g}\,f(R)
\ ,
\label{SfR}
\ee
with~\cite{zerbini,lust}
\be
f(R)=\gamma\,\lp^2\,R^2
\ ,
\label{fR2}
\ee
where $\gamma$ is a dimensionless constant.
We recall that the equation of motion following from Eq.~\eqref{SfR} for a
spatially flat FRLW metric,
\be
\d s^2
=
-\d t^2
+a^2(t)\left(\d r^2+ r^2\,\d\Omega^2\right)
\ ,
\ee
is given by~\cite{starobinsky,zerbini}
\be
6\,f'(R)\,H^2
=
R\,f'(R)-f(R)-6\,H\,\dot R\,f''(R)
\, ,
\label{eomR2}
\ee
where primes denote derivatives with respect to $R$ and dots derivatives with respect to 
the cosmic time $t$.
In particular, from Eq.~\eqref{fR2}, one obtains
\be
12\,R\,H^2
=
R^2
-12\,H\,\dot R
\ ,
\label{eom2}
\ee
and, for de~Sitter with $a(t)=e^{\sqrt{\Lambda/3}\,t}$, one has
\be
R
=6\left(H^2+\frac{\ddot a}{a}\right)
=6\left(H^2+\frac{\Lambda}{3}\right)
\ ,
\ee
and
\be
\dot R
=
6\,\left(2\,H\,\dot H+\frac{\dddot a\,a-\ddot a\,\dot a}{a^2}\right)
=
12\,H\,\dot H
\ .
\ee
By replacing the above expressions into Eq.~\eqref{eom2},
we simply obtain
\be
H^2
=
\frac{\Lambda}{3}
-\frac{4\,H^2\,\dot H}{H^2+\Lambda/3}
\ ,
\label{H2dotH}
\ee
which is solved by 
\be
H^2_\Lambda
=
{\Lambda}/{3}
\ ,
\ee
and $R=4\,\Lambda$ as expected.
\par
Upon comparing with the corpuscular description, we can therefore say that, up to a common
numerical factor, 
\be
\gn\,U_{\rm N}
\simeq
-L_\Lambda^3\,H^2_\Lambda
=
-L_\Lambda
\ ,
\label{n}
\ee
and
\be
\gn\,U_{\rm PN}
\simeq
L_\Lambda^3
\left({\Lambda}/{3}\right)
=
L_\Lambda
\ ,
\label{pn}
\ee
where we recall that $U_{\rm N}$ and $U_{\rm PN}$ follow from integrating
over the Hubble volume.
This will be our starting point to build a connection between the corpuscular model and
Starobinsky's inflation~\cite{starobinskyI}.
\section{Slow-roll Inflation}
\label{sez:infla}
The ``post-Newtonian'' analysis of the graviton condensate has shown one can have an
eternally inflating universe without the need of vacuum (matter) energy.
Of course, one next needs a source that drives the universe out of inflation.
Unlike the analysis in Ref.~\cite{cko}, this contribution may just be a
small perturbation with respect to the post-Newtonian term~\eqref{UPN0} which
breaks the balance with the Newtonian term~\eqref{UN0}.
\subsection{Starobinsky model}
\label{sub}
By means of a conformal transformation given by~\cite{DeFelice:2010aj}
\be
\tilde{g}_{\mu \nu}
=
f'(R)\,
g_{\mu \nu}
\ ,
\ee
we can rewrite the action~\eqref{SfR} in the Einstein frame as 
\be
\label{einstein-frame}
\tilde S
 \!=\!\!
\int \! d^4 x \sqrt{- \tilde g}
\left[\frac{\tilde{R}}{16 \,\pi\,\gn}
-\frac{1}{2} \tilde{g}^{\mu \nu} \partial _\mu \varphi \partial _\nu \varphi
- V(\varphi)
\right],
\qquad
\ee 
where the Ricci scalar of the original metric now appears as a new scalar field
\be 
\varphi
\equiv
\sqrt{\frac{3}{16 \,\pi\,\gn}} \, \ln f' (R)
\ee
with the potential
\be 
V(\varphi)
\equiv
\frac{f'(R(\varphi)) \, R(\varphi) - f(R(\varphi))}{16 \,\pi\,\gn\, f' (R(\varphi)) ^2}
\ .
\ee
This shows that, unless $f'$ is a constant, the original metric $g_{\mu\nu}$ contains two massless
degrees of freedom, corresponding to the helicity~2 gravitons of the metric $\tilde g_{\mu\nu}$,
and a spin~0 degree of freedom $\varphi$ associated with the trace of its Ricci tensor (see Ref.~\cite{lust}
for more details). 
\par
\begin{figure}[t!]	
\centering
\includegraphics[scale=0.22]{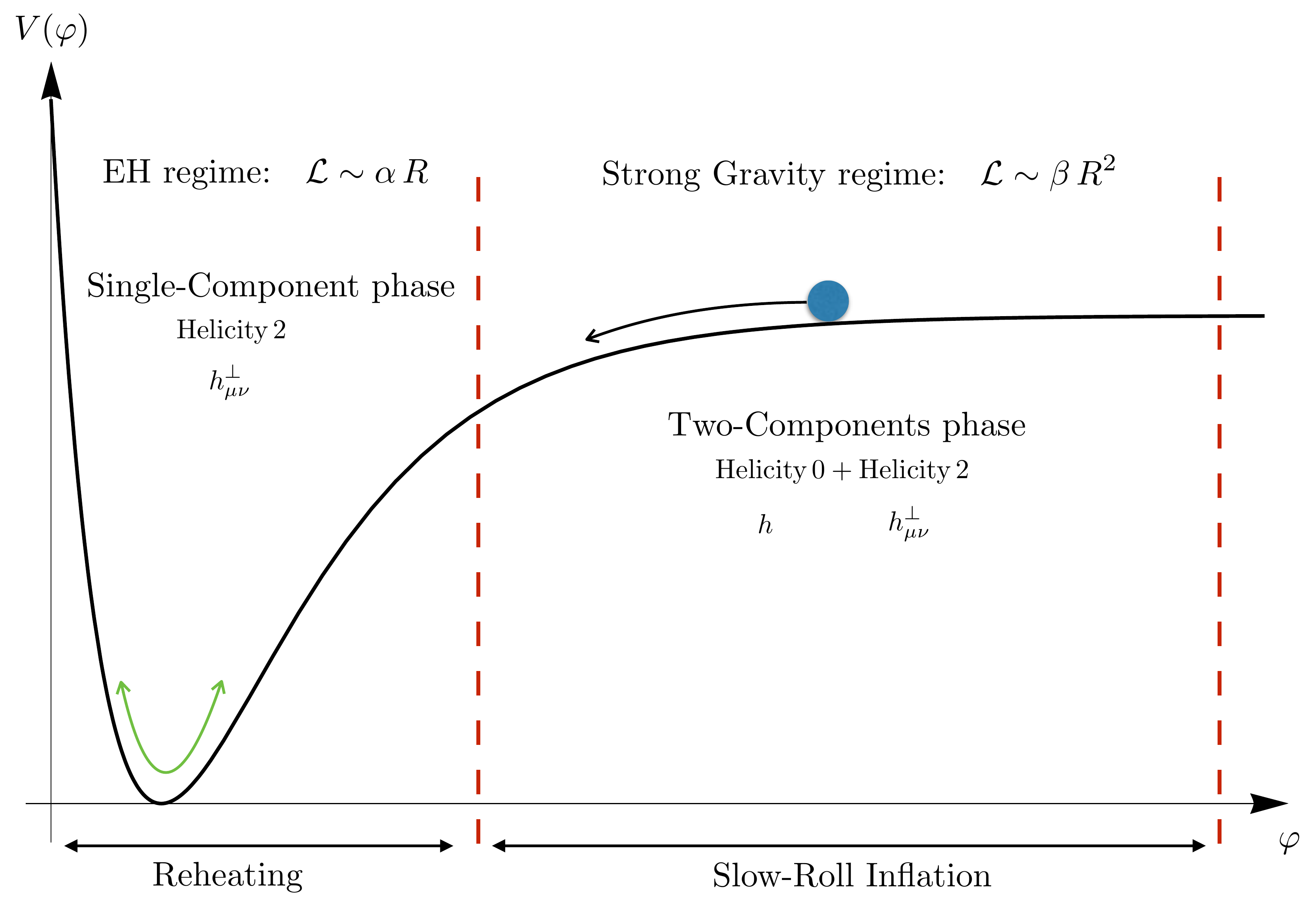}
\caption{Starobinsky's potential for the inflaton.}
\label{f1}
\end{figure}

In particular, for 
\be 
f(R)
=
\alpha\,R
+
\gamma \,\lp^2\, R^2
\ ,
\label{R+R2}
\ee
one finds
\be 
\varphi
=
\sqrt{\frac{3\,\mpl}{16 \,\pi\,\lp}}\,\ln\!\left(\alpha + 2\, \gamma \,\lp^2\, R \right)
\ ,
\ee
from which we also deduce that
\be 
R(\varphi)
= \frac{\exp\! \left( \sqrt{\frac{16 \,\pi\,\lp}{3\,\mpl}} \, \varphi \right) - \alpha}{2 \, \gamma\,\lp^2}
\ ,
\ee
and then
\be
V(\varphi;\alpha,\gamma)
=
\frac{\mpl}{64\,\pi\,\lp^3\,\gamma}
\left[ 1 - \alpha\,\exp \left( - \sqrt{\frac{16 \, \pi\,\lp}{3\,\mpl}} \, \varphi \right) \right]^2
\, , \notag
\ee
which is precisely Starobinsky's potential for the inflaton~\cite{starobinskyI}
(see Fig.~\ref{f1}).
\par
This potential has a minimum for 
\be
\varphi
= 
\frac{3\,\mpl\,\ln\alpha}{16\,\pi\,\lp}
\ ,
\ee
and
$
\lim_{\varphi\to\infty}
V(\varphi;\alpha,\gamma)
=
\frac{\mpl}{64 \, \pi \,\lp\, \gamma}
=
\lim_{\alpha\to 0}V(\varphi;\alpha,\gamma)
$.
For $\alpha=0$, one again recovers the de~Sitter space with $R=4\,\Lambda$ and
a correspondingly constant scalar field $\varphi$.
As soon as $\alpha>0$, this configuration becomes unstable, as can also be inferred
from the equation of motion~\eqref{eomR2}, which now reads
\be
6\,\frac{\alpha}{\gamma\,\lp^2}\,H^2+12\,R\,H^2
=
R^2-12\,H\,\dot R
\ .
\ee
By assuming the solution to the above equation is still of the de~Sitter form, with a
time-dependent Hubble function $H(t)\simeq H_\Lambda=\Lambda/3$, we then obtain
\be
\dot H
\simeq
-\frac{\alpha}{\gamma\,\lp^2}
\ ,
\ee
and the Hubble function is then slowly decreasing, as we expected, for $0<\alpha/\gamma\ll 1$.
In particular, the slow-roll parameter is given by
\be
\epsilon
=
-\frac{\dot H}{H^2}
\sim 
\frac{\mpl}{\lp}
\left(\frac{V'}{V}\right)^2
\ ,
\ee
and is very small along the plateau of the potential (see Fig.~\ref{f1}).
On the other hand, at the end of inflation, when the slow-roll parameter $\epsilon \sim 1$,
one infers from the CMB data~\cite{planck} that
$\gamma/\alpha\simeq 10^{8}\simeq N_\Lambda$,
and
\be
\dot H
\simeq
-L_\Lambda^{-2}
\ .
\label{dotH}
\ee 
This is precisely the relation we will now show the corpuscular description naturally yields.
\subsection{Corpuscular model}
\label{sub:corp}
In an ideal de~Sitter universe, gravitons should satisfy the balance condition~\eqref{H00}.
Let us rewrite the Hamiltonian in Eq.~\eqref{H00} as
\be
\mathcal{H}_{\rm G}^{(2)}
\simeq
\beta
\left(
U_{\rm N}
+
U_{\rm PN}
\right)
\ ,
\label{H2}
\ee
corresponding to the effective metric action~\eqref{SfR} with Eq.~\eqref{fR2}. 
Note that we introduced the dimensionless parameter $\beta>0$ of order one,
in order to keep track of this contribution.
The complete dynamics of our universe however must also include a term corresponding to the
Einstein-Hilbert action, that is
\be
\mathcal{H}_{\rm G}^{(1)}
\simeq
\alpha\,
U_{\rm N}
\ ,
\label{H1}
\ee
where $\alpha>0$ can here be viewed as the same parameter of the metric counterpart~\eqref{R+R2}.
The full energy balance is therefore given by
\be
\mathcal{H}_{\rm G}^{(1)}
+
\mathcal{H}_{\rm G}^{(2)}
\simeq
\left(\alpha+\beta\right)
U_{\rm N}
+
\beta\,U_{\rm PN}
=
0
\label{H12}
\ ,
\ee 
and, because of the term proportional to $\alpha$, we expect the expressions~\eqref{n} and \eqref{pn}
for the ideal de~Sitter condensate are no more a solution.
\par
In fact, we are interested in a stage when departures from the de~Sitter scalings are small, and
we can therefore assume that the potentials now take the slightly more general form
\be
\gn\,U_{\rm N}
\simeq
-L^3\,H^2
\label{n2}
\ee
and
\be
\gn\,U_{\rm PN}
\simeq
L^3\,L_\Lambda^{-2}
\ , 
\label{pn2}
\ee
where $L\sim L_\Lambda$ is the new Hubble radius.
Upon replacing into Eq.~\eqref{H12}, we obtain
\be
L^3
\left[
-\left(\alpha+\beta\right)
H^2
+
\beta\,L_\Lambda^{-2}
\right]
\simeq
0
\ ,
\label{H12app}
\ee
which is solved by
\be
H^2
\simeq
\frac{\beta}{\alpha+\beta}\,
\frac{1}{L_\Lambda^2}
\ .
\ee
Of course, the de~Sitter case is properly recovered when $\alpha=0$, but
$\alpha>0$ implies that $H<H_\Lambda$ as expected. 
If the system starts with $H=H_\Lambda$, the time derivative $\dot H$ must be negative
in order to ensure the constraint~\eqref{H12} holds at all times.
This can be explicitly seen by writing
\be
H
=
H_\Lambda
+
\dot H\,\delta t
\ ,
\ee
where the typical time scale $\delta t\simeq L_\Lambda$ (since gravitons of Compton length
$L_\Lambda$ cannot be sensitive to shorter times).
Eq.~\eqref{H12app} finally yields
\be
\dot H
\simeq
-\frac{\alpha}{\alpha+\beta}\,
\frac{H_\Lambda}{\delta t}
\simeq
-\frac{\alpha}{\alpha+\beta}\,
\frac{1}{L_\Lambda^2}
\ .
\label{dotHc}
\ee
We can further notice that the slow-roll parameter
\be
\epsilon
=
-\frac{\dot H}{H^2}
\simeq
\frac{\alpha}{\beta}
\label{eps}
\ee
in the corpuscular model, and one therefore obtains Eq.~\eqref{dotH} with the natural choice
$\alpha/(\alpha+\beta)\simeq 1$.
\par
Having recovered the prediction of Starobinsky's model at the end of inflation,
we can then assume that $\alpha$ and $\beta$ are proportional to the fraction of gravitons 
in the condensate whose dynamics is mostly affected by the Hamiltonian $\mathcal{H}_{\rm G}^{(1)}$
in Eq.~\eqref{H1} and $\mathcal{H}_{\rm G}^{(2)}$ in Eq.~\eqref{H2}, respectively.
At the beginning of inflation most of the $N_\Lambda$ gravitons are in the de~Sitter
condensate and just interact via the term $\mathcal{H}_{\rm G}^{(2)}\sim R^2$
(which means, $\alpha\ll 1$ and $\beta\simeq 1$), whereas
at the end of inflation all the $N_\Lambda$ gravitons interact also via the
term $\mathcal{H}_{\rm G}^{(1)}\sim R$, so that $\alpha\sim\beta\sim 1$.
In some more details, gravitons in the condensate generate the effective
Hubble expansion parameter $H\sim N_\Lambda^{-1/2}\sim L_\Lambda^{-1}$,
but they also scatter and deplete.
Their number therefore changes in time according to Eq.~(3.23) of Ref.~\cite{cko},
which we can rewrite as
\be
-\frac{\dot H}{H^2}
\simeq
\frac{\lp\,\dot N_\Lambda}{\sqrt{N_\Lambda}}
\simeq
\epsilon
\left(
1-
\frac{1}{\epsilon^{3/2}\,N_\Lambda}
\right)
\ ,
\label{dotN}
\ee
where $\epsilon\sim\alpha$ from Eq.~\eqref{eps}, the first term reproduces the background
evolution in the slow-roll approximation and the second term is due to the depletion.
It is now clear that near the end of inflation, when $\epsilon\sim 1$, the relative effect of depletion
becomes of order $N_\Lambda^{-1}$ and therefore negligibly small.
On the other hand, for~\cite{dvali}
\be
\epsilon
=
\epsilon_*
\sim
N_\Lambda^{-2/3}
\sim
\left(\frac{\lp}{L_\Lambda}\right)^{4/3}
\ ,
\label{epsC}
\ee
one obtains $\dot N_\Lambda\simeq 0$,
which can be viewed as the closest the corpuscular model can get to the pure de~Sitter space
(ideally represented by $\epsilon=\alpha=0$)~\footnote{A similar argument was already employed in
Ref.~\cite{cko} to estimate the number of $e$-foldings.}.
Equivalently, we deduce the parameter $\alpha$ will run from the minimum value of order $L_\Lambda^{-4/3}$
to the maximum of order one during the inflationary expansion.
The minimum value~\eqref{epsC} is a peculiar prediction of the corpuscular model for the slow-roll
parameter. 
\subsection{Physical outcomes}
\label{ss:phys}
As we have seen, the corpuscular model allows one to recover the background 
evolution equations of Starobinsky's model with no ambiguous coefficient at the end
of inflation, and the minimum value $\epsilon_*$ of the slow-roll parameter given in 
Eq.~\eqref{epsC} at the beginning. 
We therefore expect that the leading order phenomenology is the same as in Starobinsky's
model with initial conditions compatible with Eq.~\eqref{epsC}.
In fact, it was already shown in Ref.~\cite{cko} that the corpuscular model
correctly reproduces the behaviour determined by the given background
evolution, with corrections of order $N_\Lambda^{-1}\sim L_\Lambda^{-2}$.
\par
From the phenomenological point of view, scalar and tensor perturbations should arise from the
depletion of the background condensate.
An intriguing implication may concern the production of gravitational waves during the inflationary
process.
In fact, the dimensionless power spectrum of primordial tensor perturbations
$\mathcal{P}_{\rm T} \sim \lp^2\,H^2\sim \lp^2/L_{\Lambda}^2$~\cite{Baumann:2009ds},
will receive corrections from Eqs.~\eqref{dotN}, that is 
\be
\frac{\Delta\mathcal{P}_{\rm T}}{\mathcal{P}_{\rm T}}
\simeq
\frac{H\,\delta H}{H^2}
\sim
\frac{\dot H\, \delta t}{H}
\sim
-\epsilon
\left(
1-
\frac{1}{\epsilon^{3/2}\,N_\Lambda}
\right)
\ ,
\label{dPT}
\ee
where we again used $\delta t\sim L_\Lambda\sim H^{-1}$. 
This correction is negative and proportional to $\dot N_\Lambda/\sqrt{N_\Lambda}$:
it vanishes at the beginning of inflation, when $\epsilon\simeq \epsilon_*$,
and $\Delta\mathcal{P}_{\rm T}+\mathcal{P}_{\rm T}\simeq 0$ at the end of inflation, when $\epsilon\sim 1$.
One might be therefore tempted to relate this feature to the fact that the condensate cools down,
as the universe inflates, and the depletion of helicity~2 modes (almost) stops at the end of inflation. 
On the other hand, we should remark that the correction~\eqref{dPT} is very small
at the beginning of inflation, when $\epsilon\simeq \epsilon_*$, and it should not affect the standard
phenomenological picture in a drastic way.
For instance, the tensor-to-scalar ratio could be estimated from Eq.~(5.36) of Ref.~\cite{dvali},
and further analysed as Eq.~(4.2) in Ref.~\cite{cko}, where it was again shown that results are very
close to the ones obtained from the standard approach to cosmological perturbations.
A more quantitative analysis of the expected small corrections is left for future developments.
\par
Similar conclusions should hold for the reheating at the end of inflation, where we again 
remark that the depletion in Eq.~\eqref{dotN} leads to order $1/N_\Lambda$ corrections for the
background evolution with respect to Starobinsky's model when $\epsilon\sim 1$.
It is however known that the behaviour of the reheating phase depends strongly on the specific
particle content of the theory~\cite{Kofman}. 
Even in simple models, like chaotic inflation, one finds collective, therefore non-perturbative, 
effects, such as parametric resonance and Bose enhancement. 
The equations of motion of the fields can be rewritten, under certain approximations,
as Mathieu's differential equations, and the widths of the resonance bands then 
depend on the parameters of the theory.
For the corpuscular model, the precise set-up of such a machinery has yet to be properly
derived, and a detailed analysis is again left for future developments.
\section{Discussion and conclusions}
\label{s:conc}
We started from the simple corpuscular description of the de~Sitter universe viewed as a condensate
of $N_\Lambda$ self-interacting (scalar) gravitons of Compton wave-length equal to
$L_\Lambda\simeq H^{-1}_\Lambda$, as first proposed in Ref.~\cite{dvali}.
We then noticed that a refined, and yet equivalent description can be obtained from 
the Hamiltonian constraint with ``Newtonian'' and ``post-Newtonian'' energy terms~\cite{baryons}.
Since the de~Sitter metric is an exact solution of the modified $f(R)\simeq R^2$ theory of gravity,
we inferred that it should also be possible to reinterpret this quantum state in terms of an effective
metric theory of this form.
Moreover, this theory is equivalent to the usual Einstein-Hilbert gravity with the addition of a scalar field
(replacing the trace of the Ricci scalar), and therefore contains one more degree of freedom (of helicity~0)
than the Einstein theory (which contains two helicity~2 modes).
In the pure de~Sitter, we hence expect all degrees of freedom are in an equilibrium state solely
characterised by the length scale $L_\Lambda$. 
\par
Of course, any realistic model of inflation requires a departure from (eternal) de~Sitter, which can be
achieved by adding the Einstein-Hilbert term $R$ to the previous $f(R)\simeq R^2$ theory in the metric
description.
Since the corpuscular description of the pure Einstein gravity is just given by the ``Newtonian'' term, 
this is equivalent to introducing such an extra term that pushes the de~Sitter gravitons off equilibrium.
We have seen that this mechanism is compatible with the Hamiltonian constraint and, indeed, it appears
that the length scale $L_\Lambda$ naturally fixes the size of $\dot H$ at the end of inflation
to the one required by experimental data, as well as the value~\eqref{epsC} of the slow-roll parameter at the
beginning of inflation.
To summarise, the corpuscular model of inflation contains one scale $L_\Lambda$ from which the main
dynamical features of the inflationary background can be extracted.
\par
We conclude by mentioning that it was recently shown in Ref.~\cite{Cadoni:2017evg} how the same corpuscular
description of the quantum state of the universe can also explain the observed galaxy rotation curves without
the need of dark matter.
More detailed quantitative analysis of both inflation and dark matter phenomenology are part of
future developments. 
\acknowledgments
We would like to thank G.~Dvali for useful comments.
R.C.~and A.~Giusti~are partially supported by the INFN grant FLAG.
A.~Giugno was partially supported by the ERC Advanced Grant~339169 {\em Selfcompletion\/}.
This work has also been carried out in the framework of activities of the National Group
of Mathematical Physics (GNFM, INdAM).
\end{document}